\documentstyle[12pt]{article}
\newcommand{\beq}{\begin{equation}}
\newcommand{\eeq}{\end{equation}}

\begin{document}
\begin{titlepage}
\begin{flushright}
NBI-HE-93-50\\
August 1993\\
\end{flushright}
\vspace{0.5cm}
\begin{center}
{\large {\bf On String Tunneling in Power Law Inflationary Universes.}}\\
\vspace{1.5cm}
{\bf A.L.Larsen}
\footnote{e-mail:allarsen@nbivax.nbi.dk}\\
\vspace{0.4cm}
{\em Nordita, 17 Blegdamsvej, 2100 Copenhagen, Denmark} \\
\vspace{0.4cm}
{\bf Minos Axenides}
\footnote{e-mail:axenides@nbivax.nbi.dk}\\
\vspace{0.4cm}
{\em The Niels Bohr Institute\\
University of Copenhagen, 17 Blegdamsvej, 2100 Copenhagen, Denmark}\\
\vspace{0.4cm}
\end{center}
\vspace{6mm}
\begin{abstract}
We consider the evolution of circular string loops in power law
expanding universes represented by a spatially flat Friedman-Robertson-Walker
metric with scale factor $a(t)\propto t^{p}$, where $t$ is the cosmic time and
$p\geq 0$. Our main result is the existence of a ``magic'' power
$p_{m}= 3+2\sqrt{2}$. In spacetimes with $p<p_{m}$ a circular string
expands either forever or to a maximal radius and then
contracts until it collapses into a point (black hole). For $ p\geq
p_{m}$,
however, we find additional types of solutions. They include
configurations which contract from a positive initial radius to a minimal
one and then expand forever. Their existence we interpret as an
indication for the presence of a finite potential barrier.
Equivalently the new solutions signal string
nucleation and tunneling, phenomena recently shown to occur
in de Sitter space.

\end{abstract}
\end{titlepage}
\newpage

\section{Introduction}
Inflation \cite{Linde} is a period of rapid expansion in the early
history of the universe. Through its different realizations (power
law, intermediate, exponential etc.) many cosmological paradoxes find
a simultaneous resolution. Topological defects such as cosmic strings,
monopoles and domain walls are typically generated at cosmic phase
transitions through the Kibble mechanism \cite{Kolb}. It has been a
theoretical prejudice that an unwanted overabundance of such objects
in the observable universe can be eliminated by an inflationary phase
following such phase transitions. More specifically the cosmic string
scenario for the generation of the large scale structure was considered
to be hardly compatible with inflation in a single theoretical
framework \cite{Bob}.

Recently in a very interesting paper \cite{Basu1} Basu, Guth and Vilenkin
took a step towards resolving the conflict by demonstrating
the spontaneous nucleation of cosmic strings in de Sitter
space. Using the static parametrization they found that the classical
evolution of a circular string is determined by a simple potential
barrier and that strings can nucleate by a quantum mechanical tunneling
through the barrier. The tunneling amplitude that determines the
nucleation rate could finally be calculated
using either the WKB approximation or the instanton method.
String nucleation of ref.(4) has been further investigated in
refs.(5-9). The case of charged string tunneling processes
in de Sitter space has also appeared in the literature \cite{Nielsen}.

The purpose of the present work is to consider the possibility of string
nucleation in power law inflationary universes \cite{PL}.
The first problem we meet
is that such spacetimes do not have a static parametrization. Therefore
we cannot expect that the evolution of classical circular string loops
is determined by a simple one dimensional Hamiltonian with a simple
potential, from which the tunneling amplitude can be calculated
explicitly, as was the case in de Sitter space \cite{Basu1}.
Secondly, the equations of motion for
the circular strings are not even integrable in the cases under
consideration here. We will therefore use more indirect analytical and
numerical methods, and will restrict ourselves to formulating only
necessary and sufficient conditions on time dependent spacetimes for
string nucleation to happen, and moreover argue for its occurence in
the case to be considered.

Our starting point is that string nucleation of circular loops is
intimately connected to the phenomenon of string tunneling
\cite{Basu1}. We therefore expect that string nucleation may play an
important role in cosmological spacetimes that admit circular string
configurations with the property :
\begin{equation}
\exists\; t_{0} :\;\; f(t) \geq f(t_{0}) > 0 ,\;\;\;  \frac{df}{dt}(t_{0})=0
\end{equation}
where $f$ is the physical radius of the loop and $t$ is the cosmic time.
Loosely speaking eq.$(1.1)$ expresses that there are circular
string configurations which are ``energetically'' forbidden to collapse,
i.e. if they contract they will sooner or later hit a ``barrier'' (at
$t=t_{0}$). The existence of such a barrier on the other hand means that
tunneling effects may be relevant and that circular strings may
nucleate with finite radius. We can also of course imagine strings
collapsing by tunneling the other way through the barrier.

In the
present paper we will consider time dependent spacetimes and therefore
the condition $(1.1)$ must hold only locally around $t_{0}$, i.e. for :
\begin{equation}
t \in \left [ t_{0} - \Delta t_{1}, t_{0} + \Delta t_{2} \right ]
\end{equation}
for some positive $\Delta t_{1,2}$. Loosely speaking we may say that a
barrier that prevents a string from collapsing at $t=t_{0}$ will change
in time, and in some cases may only exist for a finite time during the
evolution of the universe. This means that a nucleated string which is
created by tunneling through a barrier, in a later moment of the
cosmic expansion (where the original potential barrier has disappeared
again) can collapse by continuously and classically shrinking to a point
\cite{Hawking}. We believe the above picture to be realistic and
physical as our universe went through a de Sitter like phase in its
early moments before it entered the radiation dominated era. In the
latter phase the string evolution equations indicate an unconditional
string collapse. This is also the physical picture that was recently
adopted by Garriga and Vilenkin\cite{G3} where de Sitter spacetime is
patched next to a Minkowski one as an approximation to the radiation
dominated universe. It should be stressed that in what follows we are not
directly studying the nucleated strings. Instead we use ``test strings''
with all kinds of initial values of radii and velocities in order to
explore the structure of the interaction between the string and the
underlying expanding universe. As explained above, this
approach provides information about the possible existence of tunneling
phenomena and string nucleation processes.

The paper is organized as follows. In section $2$ we derive the
equations of motion that determine the classical evolution of circular
strings in a spatially flat Friedman-Robertson-Walker(FRW)
universe \cite{Akdeniz,Vilenkin,Li,Vega}, in a gauge where the string time is
identified with the cosmic time. In section $3$ we analyze the equations
of motion in power law expanding universes and find a ``magic'' power
related to tunneling phenomena. We also present numerical results that
confirm the significance of this magic power and we discuss some of the
string trajectories. Finally in section $4$
we present our conclusions.
\section{The equations of motion}
\setcounter{equation}{0}
We consider the equations of motion for a circular string in a spatially
flat FRW spacetime, in the special case of the scale factor
being a power of the cosmic time. We therefore take the line element in
the form:
\begin{equation}
ds^{2} = -dt^{2} + a(t)^{2}\left(
dr^{2}+r^{2}d\theta^{2}+r^{2}\sin^{2}\theta d\phi^2\right)
\end{equation}
and we will eventually take
\begin{equation}
a(t) = a_{p} t^{p}\;\; ; \;\;\;\;p\geq 0
\end{equation}
where $a_{p}$ is a dimensionful constant which is included to ensure that
$a(t)$ is dimensionless. Note that the family of spacetimes parametrized
in $(2.2)$ includes Minkowski space $(p=0)$, a radiation dominated
universe $(p=1/2)$, a matter dominated universe $(p=2/3)$, a linearly
expanding universe $(p=1)$ and power law inflationary universes
$(p>1)$.

Before we embed a circular string let us recall a few facts about these
spacetimes. It is the existence of either particle horizons or event
horizons\cite{GR}. For $1>p\geq 0$ there is a particle horizon with a physical
radius given by :
\begin{equation}
f_{PH}(t) = a(t)\;\; r_{PH}(t) \equiv a(t)\;\int_{0}^{t} \frac{ds}{a(s)}
\;=\;\;\frac{t}{1-p}
\end{equation}
while for $p>1$ there is an event horizon with a physical radius given by:
\begin{equation}
f_{EH}\;=\;a(t)\;r_{EH}(t)\;\equiv\;a(t)\;\int_{t}^{\infty}\;
\frac{ds}{a(s)}\;=\;\frac{t}{p-1}
\end{equation}
Both types of horizon will be essential later on in our analysis of the
evolution of circular strings. It is also worth keeping in mind the
expressions for the nonvanishing components of the Riemann tensor. In
cartesian coordinates we have that:
\begin{eqnarray}
R_{xtxt}\hspace*{-2mm}&=&\hspace*{-2mm}R_{ytyt} =
R_{ztzt} =-a(t)\;\frac{d^{2}a(t)}{dt^{2}}\;=\;
-a_{p}^{2}p(p-1)t^{2(p-1)} \nonumber \\
R_{xyxy}\hspace*{-2mm}&=&\hspace{-2mm}R_{xzxz} =
R_{yzyz} = a(t)^{2}\left(\frac{da(t)}{dt}
\right)^{2} = a_{p}^{4}p^{2}t^{2(2p-1)}
\end{eqnarray}
So spacetimes $(2.1-2.2)$ are plagued with a spacetime singularity at $t=0$
unless either $p=0$ or $p\geq 1$.

Let us now consider a classical string
described by the Nambu-Goto action:
\begin{equation}
{\cal L} \sim \sqrt {-\det G_{\alpha\beta}}
\end{equation}
where $G_{\alpha\beta}$ is the induced metric on the world-sheet
\begin{equation}
G_{\alpha\beta}\;=\;g_{\mu\nu}\;X^{\mu}_{,\alpha}\;X^{\nu}_{,\beta}
\end{equation}
with  $X^{\mu}=(t,r,\theta,\phi)$ and
$g_{\mu\nu}=diag (-1,a^{2},a^{2}r^2,a^{2}r^2\sin^2\theta)$. For completeness we
have kept
the scale-factor arbitrary. A circular string is most easily obtained
by the following combined gauge choice and {\it Ansatz}:
\begin{equation}
t=\tau,\;\;\;r=r(t),\;\;\;\theta=\pi/2,\;\;\;\phi=\sigma
\end{equation}
where $(\tau,\sigma)$ are the two internal world-sheet coordinates. The
equations of motion obtained from the Lagrangian $(2.6)$ and the ansatz
$(2.8)$ are:

\[\frac{d}{dt}\left[\frac{a^{2}r^{2}}
{\sqrt{a^{2}r^{2}(1-a^{2}\dot{r}^{2})}}\right]+
\frac{\dot{a}ar^{2}(2a^{2}\dot{r}^{2}-1)}{\sqrt{a^{2}
r^{2}(1-a^{2}\dot{r}^{2})}}=0\]
\begin{equation}
\frac{d}{dt}\left[\frac{a^{4}r^{2}\dot{r}}
{\sqrt{a^{2}r^{2}(1-a^{2}\dot{r}^{2})}}
\right]
+\frac{a^{2}r(a^{2}\dot{r}^{2}-1)}{\sqrt{a^{2}r^{2}(1-a^{2}\dot{r}^{2})}}=0
\end{equation}
where the dot denotes differentiation with respect to $t$. This looks
quite complicated but after a little algebra one finds \cite{Akdeniz,Li}
that both equations reduce to:
\begin{equation}
%% FOLLOWING LINE CANNOT BE BROKEN BEFORE 80 CHAR
a^{2}r\ddot{r}\;-\;2r\dot{a}a^{3}\dot{r}^{3}\;-\;a^{2}\dot{r}^{2}\;+\;3a\dot{a}r
\dot{r}\;+\;1=\;0
\end{equation}
For our purposes it is convenient to introduce the physical string
radius:
\begin{equation}
f\;=\;a\;r
\end{equation}
and the Hubble function:
\begin{equation}
H\;=\;\frac{\dot{a}}{a}
\end{equation}
Equation $(2.10)$ becomes:
\begin{eqnarray}
\ddot{f}f\;-\;2Hf\dot{f}^{3\hspace*{-1mm}}&+&\hspace*{-1mm}(6H^{2}f^{2}-1)
\dot{f}^{2}\;+\;3Hf(1-2H^{2}f^{2})\dot{f}\nonumber\\
\hspace*{-1mm}&-&\hspace*{-1mm}f^{2}\dot{H}\;+\;2H^{4}
f^{4}\;-3H^{2}f^{2}\;+\;1=\;0
\end{eqnarray}
The above equation gives the physical string radius $f$ as a function of the
cosmic time in a spatially flat FRW spacetime with a Hubble function H.
It's exact analytical solution does not seem to be available,
but it is very suitable for a numerical investigation. With regard to
the boundary conditions we should keep in mind that a timelike string is
obtained provided:
\begin{equation}
G_{00}=\;-1+a^{2}\dot{r}^{2}\;=\;-1+(\dot{f}\;-\;H\;f)^{2}\;\;<\;\;0
\end{equation}
In the rest of our investigation we will look for configurations that
satisfy the above inequality. For the special case of
 $a(t)=a_{p}t^{p}$ and $H(t)=p/t$ eq.$(2.13)$ takes the form:
\begin{equation}
\ddot{f}f\;-\frac{2p}{t}f\dot{f}^{3} + (\frac{6p^{2}}{t^{2}}f^{2}-1)\dot{f}^{2}
+
\frac{3p}{t}(1-\frac{2p^{2}}{t^{2}}f^{2})\dot{f}f +
\frac{2p^{4}}{t^{4}}f^{4}+p(1-3p)\frac{f^{2}}{t^{2}} + 1=0
\end{equation}
Both eqs.$(2.13, 2.15)$ will be studied in more detail in the next
section.

\section{The magic power and numerical solutions}
\setcounter{equation}{0}
Let us assume that for some scalefactor $a(t)$ we find a string
configuration that satisfies conditions  $(1.1)$
and $(1.2)$ at some fixed $t_{0}$. By a Taylor expansion of $f(t)$ around
$t_{0}$ we get that $\ddot{f}(t_{0})\geq 0$. The equation of motion
$(2.13)$ for $f(t)$ gives at $t=t_{0}$ :
\begin{equation}
\dot{H}(t_{0})\;\geq\;\frac{1}{f(t_{0})^{2}}+2H(t_{0})^{4}f(t_{0})^{2} -
3 H(t_{0})^{2}
\end{equation}
On the other hand condition $(2.14)$ gives for $t=t_{0}$ :
\begin{equation}
\frac{1}{H(t_{0})}\;\;>\;\;f(t_{0})\;\;>\;\;0
\end{equation}
and by combining $(3.1)$ with $(3.2)$ we find:
\begin{equation}
\dot{H}(t_{0})\;\;\geq\;\;(2\sqrt{2}-3)H(t_{0})^{2}
\end{equation}
According to the discussion of the interpretation of $t_{0}$ (see the
introduction) this inequality gives a {\em necessary} condition for the
existence of string loop tunneling
and nucleation. We may note in passing that $(3.3)$ is trivially
fulfilled for the de Sitter spacetime $(a\sim e^{t},H=const)$.
A remarkable fact
about power law expanding universes is that $\dot{H}$ is proportional to
$H^{2}$ so that $t_{0}$ drops out of $(3.3)$. In fact from $(2.2)$ we
get:
\begin{equation}
p\;\;\geq \;3+2\sqrt{2}\;\;\equiv \;p_{m}
\end{equation}
which is exactly the ``magic''power advertised in the abstract. If
$p\;<\;p_{m}$ there can be no string configuration  fulfilling
conditions $(1.1-1.2)$. This implies that a circular string will either
expand monotonically for ever, contract monotonically until it collapses,
or expand monotonically to a maximal radius
$f(\acute{t})$ (where$\dot{f}(\acute{t})=0,\;\ddot
{f}(\acute{t})<0)$ and then contract
monotonically until it collapses. Examples of such configurations
obtained by numerically integrating equation $(2.15)$ are shown in
figs.$1$ and $2$. It should be remarked that from a mathematical point
of view the collapsing strings can be continued below $f=0$ and
then appear as oscillating. We, however, take the cosmic string point
of view where the
evolution of the string stops the first time the physical radius reaches
zero from above and the string has collapsed into a black
hole \cite{Hawking}. Let us
now consider spacetimes where $p\geq p_{m}$, i.e. for which inequality
$(3.4)$ is satisfied. In this case we expect to find configurations
contracting to a minimal radius, subsequently expanding. By numerically
integrating eq.$(2.15)$ we have indeed found such solutions for all
$p\geq p_{m}$. We present two such examples in figures $3$ and $4$. Note
that after bouncing at the barrier the string, depending on its initial
radius and velocity, can either continue to expand forever (fig.$3$) or
actually collapses ultimately anyway (fig.$4$). In addition we found
solutions of the form presented in figs.$1$ and $2$. In summary the
solutions for $p\geq p_{m}$ include strings bouncing from below (smaller
radius), strings bouncing from above (larger radius) and strings
monotonically expanding from $f=0$ to $f\rightarrow\infty$.
This fact indicates that the classical evolution of strings is
determined by a finite ``potential barrier'' and we therefore expect that
string tunneling and nucleation take place for spacetimes with
$p\geq p_{m}$. We should stress once more however, that we are
considering time-dependent spacetimes that do not admit static
parametrizations. As a consequence the apparent
potential barriers are to be interpreted as being similarly time
dependent, in contrast to the case of de Sitter spacetime \cite{Basu1}.
\subsection*{Conclusion}
In the present work we explored the possibility that quantum nucleation
of topological defects such as cosmic strings, is not a property
of the de Sitter spacetime alone. To that end we studied the circular
string evolution equations in spacetimes with variable power law
expansion rate. Surprisingly, in addition to purely expanding and
collapsing solutions, we found a new type of solutions. They occur for
spacetimes with a power greater than $3+2\sqrt{2}$ which we interpreted
as evidence for the presence of string nucleation. Quantum tunneling of
circular strings appears to occur in almost all types of generalized
inflationary cosmologies. A full classification of the different types of
circular string evolution in such spacetimes will be the subject of
future work.

\newpage
\subsection*{Figure Captions}
\vskip 6pt
Fig.1. Examples of string trajectories for $p=1/2$ (radiation dominated
universe). The particle horizon $f_{PH}=2 t$ can only be passed from the
outside, and soon after passing it the circular strings collapse \cite{Li}.
\vskip 12pt
\hspace*{-6mm}Fig.2. Examples of string trajectories for $p=2$. The presence
of an event horizon $f_{EH}=t$ guarantees the existence of non-collapsing
solutions: Once outside, the string expands towards infinity. We however
also find collapsing solutions in this case. They are of course always
inside the event horizon.
\vskip 12pt
\hspace*{-6mm}Fig.3. Example of a string
trajectory for $p=25$, i.e. $p> p_m$.
The string contracts to a minimal size, then expands through the event horizon
$f_{EH}=t/24$ and continues towards infinity.
\vskip 12pt
\hspace*{-6mm}Fig.4. Another example of a string trajectory for $p=25$. The
string contracts to a local minimal size, then expands to a maximal size
inside the event horizon and finally collapses to a point.
\end{document}